\documentclass[letterpaper, 10 pt, conference]{ieeeconf}  %

\usepackage{cite}
\usepackage{amsmath,amssymb,amsfonts}
\usepackage{algorithmic}
\usepackage{graphicx}
\usepackage{textcomp}
\usepackage{bm}
\usepackage{subcaption}

\usepackage{comment}
\usepackage{hyperref}
\usepackage{siunitx}
\usepackage[capitalise]{cleveref}
\usepackage{soul}
\usepackage{amsthm}

\usepackage{empheq}

\DeclareMathAlphabet\mathbfcal{OMS}{cmsy}{b}{n}

\newcommand*{\genbf}[1]{\ifmmode\mathbf{#1}\else\textbf{#1}\fi}
\newcommand{\norm}[1]{\left\lVert#1\right\rVert}

\newcommand{\xrefb}[0]{\mathbf{x_{ref}}}
\newcommand{\Xrefb}[0]{\mathbf{X_{ref}}}
\newcommand{\Xrefba}[0]{\mathbf{X_{ref,a}}}

\newcommand{\wb}[0]{\mathbf{w}}

\newcommand{\Psie}[0]{\bm{\Psi}^{\boldsymbol{\eta}}}
\newcommand{\Phieno}[0]{\bm{\Upsilon}^{\boldsymbol{\eta}}}
\newcommand{\Phiuno}[0]{\bm{\Upsilon}^{\mathbf{u}}}

\newcommand{\Phie}[2]{\bm{\Phi}^{\boldsymbol{\eta}}_{\mathbf{#1},\mathbf{#2}}}
\newcommand{\Phier}[2]{\bm{\Phi}^{\mathbf{e}}_{\mathbf{#1},\mathbf{#2}}}

\newcommand{\Phiu}[2]{\bm{\Phi}^{\mathbf{u}}_{\mathbf{#1},\mathbf{#2}}}

\newcommand{\Psiu}[0]{\bm{\Psi}^{\mathbf{u}}}

\newcommand{\Emme}[0]{\boldsymbol{\mathcal{M}}}

\newcommand{\bmx}{\bold{x}}
\newcommand{\bmu}{\bold{u}}
\newcommand{\bme}{\bold{e}}
\newcommand{\bmF}{\bold{F}}
\newcommand{\bmK}{\bold{K}}
\newcommand{\bmcalM}{\boldsymbol{\mathcal{M}}}

\newcommand{\xb}[0]{\mathbf{x}}

\newcommand{\eb}[0]{\mathbf{e}}
\newcommand{\ub}[0]{\mathbf{u}}

\newcommand{\vb}[0]{\mathbf{v}}

\newcommand{\etab}[0]{\boldsymbol{\eta}}
\newcommand{\bmC}{\bold{C}}

\newcommand{\hatbf}[1]{\widehat{\mathbf{#1}}}

\newcommand{\Fb}[0]{\mathbf{F}}

\usepackage{tikz}
\usetikzlibrary{arrows}

\newcommand{\Effe}[0]{\mathbfcal{F}}

\newtheorem{theorem}{Theorem}

\newtheorem{remark}{Remark}
\newtheorem{definition}{Definition}

\newtheorem{problem}{Problem}

  {\end{proof}}   % Qué hacer al terminal
\IEEEoverridecommandlockouts
\title{Boosting the transient performance of reference tracking controllers with neural networks}
\author{Nicolas Kirsch, Leonardo Massai and Giancarlo Ferrari-Trecate% <-this % stops a space
\thanks{This research has been supported by the Swiss National Science Foundation under the NCCR Automation (grant agreement 51NF40\_180545) and the NECON project (grant number 200021-219431).}% <-this % stops a space
\thanks{The authors are with the Institute of Mechanical Engineering, Ecole Polytechnique Fédérale de Lausanne (EPFL), CH-1015 Lausanne, Switzerland. (email: \tt\small {\{nicolas.kirsch@epfl.ch) } }%
}

\begin{document}

\maketitle
\begin{abstract}
Reference tracking is a fundamental goal in many control systems, particularly those with complex nonlinear dynamics. While traditional control strategies can achieve steady-state accuracy, they often fall short in explicitly optimizing transient performance. Neural network controllers have emerged as a flexible solution to handle nonlinearities and disturbances, yet they typically lack formal guarantees on closed-loop stability and performance. To bridge this gap, the recently proposed Performance Boosting (PB) framework offers a principled way to optimize generic transient costs while preserving the $\mathcal{L}_p$-stability of nonlinear systems.\

In this work, we extend the PB framework to tackle reference tracking problems. First, we characterize the complete set of nonlinear controllers that retain the tracking properties of a given baseline reference-tracking controller. Next, we show how to optimize transient costs while searching within subsets of tracking controllers that incorporate expressive neural network models. We also analyze the robustness of the proposed method under uncertainties in the system dynamics. Finally, numerical experiments on a robotic system illustrate the performance gains of our approach compared to the standard PB framework.

\end{abstract}

\section{Introduction}

Reference tracking is a fundamental objective in many control systems, playing a key role in diverse applications such as power systems \cite{reftrack_grids}, robotics \cite{reftrack_robotics}, and aerospace \cite{reftrack_aerospace}. In these fields, maintaining accurate tracking of desired setpoints is essential for system performance and reliability. Standard reference tracking controllers, such as PID, have been widely used and are generally successful in ensuring steady-state accuracy. However, these methods often do not explicitly optimize a performance metric, which can result in suboptimal behavior, especially in the presence of nonlinearities and disturbances. Beyond steady-state tracking accuracy, many applications require optimizing additional performance criteria, such as minimizing energy consumption, reducing transient overshoot, or improving disturbance rejection. These objectives are essential for ensuring efficient and safe operation but are not inherently addressed by conventional control methods. As a result, standard controllers may struggle to meet performance requirements in complex environments, particularly when dealing with nonlinear systems.

Various approaches have been explored to enhance reference tracking in nonlinear systems. Model predictive control (MPC) is a widely used method, offering the ability to track time-varying references while incorporating secondary, potentially nonlinear objectives through constraints or the loss function \cite{mpc_rt, mpc_rt3, CISNEROS201711601}. Some MPC formulations also provide various forms of closed-loop guarantees \cite{mpc_rt2, mpc_linearize}. However, the deployment of MPC policies typically requires solving complex optimization problems in real time. This can be computationally overwhelming when dealing with highly nonlinear models and cost functions \cite{rawlings2017model}.

An alternative framework is provided by adaptive control, where the controller is adjusted in real-time based on the system's performance \cite{Nguyen2018}. Some adaptive approaches also allow enforcing constraint satisfaction using barrier functions, either on output \cite{output_barrier, output_adaptive} or state variables \cite{state_adaptive}. While these methods are promising, closed-loop guarantees are usually restricted to stability and constraint satisfaction and often apply only locally.

Fuzzy logic controllers \cite{fuzzy_1, fuzzy_2} have been also used for reference tracking, leveraging their ability to approximate uncertain nonlinear dynamics, but they do not explicitly minimize loss functions encoding for additional performance objectives.

Similarly, neural networks have been applied to reference tracking, either as observers for state estimation within tracking controller \cite{PARK2017353, gao2019long}, or as optimizers of reference tracking metrics, like settling time and accuracy. However, these approaches do not provide closed-loop guarantees \cite{zhang2022neural}.

A neural network control scheme for linear systems that provides tracking guarantees by enforcing Lyapunov-like inequalities during optimization has been proposed in \cite{pauli2021offset}. For stabilization at the origin, other works have used constrained optimization to design neural network controllers with formal stability guarantees \cite{berkenkamp2017safe}. However, these approaches impose conservative stability constraints, limiting admissible policies. Furthermore, enforcing conditions like linear matrix inequalities quickly becomes a computational bottleneck in large-scale applications.

The authors of \cite{10633771} introduced the Performance Boosting (PB) framework, which leverages the flexibility of neural network controllers, while preserving closed-loop guarantees in a state-feedback setup. PB control design amounts to unconstrained optimization over state-feedback policies characterized by specific classes of neural networks, that inherently preserve the $\mathcal{L}_p$ stability property of a nonlinear system. 
This framework, which effectively decouples performance optimization from stability constraints, relies on a preliminary result showing that all and only stability-preserving controllers for a nonlinear system can be built using an Internal Model Control (IMC) scheme including an $\mathcal{L}_p$-stable operator $\Emme$. This operator can then be freely selected to optimize the desired performance metric. To avoid solving an infinite-dimensional optimization problem, in practice, $\Emme$ is chosen within a class of neural networks depending on a finite number of parameters and describing  $\mathcal{L}_p$-stable operators as those proposed in \cite{REN,forgione2021dynonet,zakwan2024neural,wang2023direct}.   In \cite{10633771}, it is also shown that under a finite $\mathcal{L}_p$-gain assumption on the model mismatch, stability can always be preserved by embedding a nominal system model in the controller and optimizing over operators $\Emme$ with a sufficiently small $\mathcal{L}_p$-gain.

%\cite{10633771} indeed shows that an internal model control (IMC) \cite{GarciaMorari1982} architecture allows to characterize, without conservatism, the class of \emph{all and only stability-preserving controllers} where the only free parameter $\Emme$ is an $\mathcal{L}_p$ operator. As long as this condition is satisfied, $\Emme$ can be parametrized in any specific way which improves the performance of the closed-loop. A possible structure for this operator is as a recurrent equilibrium network (REN) \cite{REN}, a specific type of recurrent neural network which can be freely parametrized while providing an $\mathcal{L}_p$ operator. By using RENs, it is possible to design the controller using unconstrained methods to optimize any cost, while always ensuring the resulting operator is $\mathcal{L}_p$. The closed-loop system will be $\mathcal{L}_p$ stable even if the optimization is halted prematurely, as this property is independent of the optimization process. Any type of nonlinear objective can be achieved by appropriately designing the loss function that the REN minimizes.

%They also shown that under a finite gain assumption on the model mismatch, stability can always be preserved by embedding a nominal system model and optimizing over nonlinear controllers with a sufficiently reduced $\mathcal{L}_p$ gain on the free parameter.

\subsection*{Contributions} We introduce a new neural network control method for reference tracking tasks, which we call rPB (reference PB). We derive a complete parametrization of \emph{all and only reference tracking controllers} in terms of a free operator satisfying mild conditions on its input-to-output behavior. This characterization is independent of the specific signal to be tracked, allowing a single training process to generalize across a wide range of references. Any nonlinear performance metric can be optimized in an unconstrained manner within the rPB framework. Additionally, we establish robustness guarantees by proving that reference tracking guarantees are preserved under imperfect model knowledge, provided the model mismatch is an incrementally-finite-gain $\ell_p$-stable (denoted i.f.g $\ell_p$-stable) operator.

To illustrate the effectiveness of rPB, we apply it to a simulated robotic system. We highlight its advantage over standard PB by demonstrating its ability to generalize across multiple reference targets. We then showcase its capacity to generate optimal trajectories in a highly nonlinear, non-convex setting, showing its potential for complex real-world applications.

\subsection*{Notation}
The set of all sequences $\mathbf{x} = (x_0,x_1,x_2,\ldots)$, where $x_t \in \mathbb{R}^n$, $t\in \mathbb{N}$ , is denoted as $\ell^n$. 
Moreover,  $\mathbf{x}$ belongs to $\ell_p^n \subset \ell^n$ with $p \in \mathbb{N}$ if $\norm{\mathbf{x}}_p = \left(\sum_{t=0}^\infty |x_t|^p\right)^{\frac{1}{p}} < \infty$, where $|\cdot|$ denotes any vector norm. 
We say that $\xb \in \ell^n_\infty$ if $\operatorname{sup}_{t}|x_t|< \infty$. 
When clear from the context, we omit the superscript $n$ from $\ell^n$ and $\ell^n_p$. 
An operator $\mathbf{A}:\ell^n \rightarrow \ell^m$ is said to be \emph{causal} if $\mathbf{A}(\mathbf{x}) = (A_0(x_0),A_1(x_{1:0}),\ldots,A_t(x_{t:0}),\ldots)$, and $\mathbf{A}$ is said to be $\ell_p$-stable if it is \emph{causal} and $\mathbf{A}(\mathbf{w}) \in \ell_{p}^m$ for all $\mathbf{w} \in \ell_{p}^n$. Equivalently, we  write $\mathbf{A} \in \mathcal{L}_{p}$. 
We say that an $\mathcal{L}_p$ operator $\mathbf{A}:\mathbf{w}\mapsto \mathbf{u}$  has finite $\mathcal{L}_p$-gain $\gamma(\mathbf{A})>0$ if  $\|\mathbf{u}\|_p\leq \gamma(\mathbf{A})\|\mathbf{w}\|_p$, for all $\mathbf{w}\in\ell_p^n$. 
Similarly, we say that an operator \(\mathbf{A}:\mathbf{w} \mapsto \mathbf{u}\) is i.f.g $\ell_p$-stable and has finite incremental \(\mathcal{L}_p\)-gain  \(\alpha(\mathbf{A}) > 0\) if for any \(\mathbf{w}_1, \mathbf{w}_2 \in \ell_p^n\), the output difference satisfies $\|\mathbf{u}_1 - \mathbf{u}_2\|_p \leq \alpha(\mathbf{A}) \|\mathbf{w}_1 - \mathbf{w}_2\|_p$.

\section{Problem formulation}

We study nonlinear discrete-time time-varying systems augmented by a base controller achieving reference tracking. For example, one can consider the tracking of a constant but a priori unknown reference using an integral control action. At time $t$, the plant state is $x_t \in \mathbb{R}^n$ and the base controller state is $v_t \in \mathbb{R}^v$. The total augmented state of the base system is $\eta_t = (x_t^\top,v_t^\top)^\top \in \mathbb{R}^{q}$, with $q = n + v$. The set point is $x_{ref,t} \in \mathbb{R}^n $. The dynamics are given by: 
\begin{equation}\label{eq:system_state}
    \eta_t = f_t(\eta_{t-1:0},u_{t-1:0},x_{ref,t-1:0})+ w_t, ~~~t= 1,2,\ldots\,,
\end{equation}
where $u_t \in \mathbb{R}^m$ is an control input affecting the base system, $w_t \in \mathbb{R}^q$ is an unknown process noise with $ w_0 =(x_0^\top,0_v^\top)^\top $, and $f_0=0$. The noise influences the system states but not the base controller states, so  $w_t = ({w_{x,t}}^\top,0_v^\top)^\top$ with $ w_{x,t} \in \mathbb{R}^n$ for all $t$. \footnote{In spite of the dependence of $f$ and $w$ on the index $t$ (instead of $t-1$), \eqref{eq:system_state} defines a standard nonlinear system where the initial state is assigned through $w_0$} We consider disturbances with support $\mathcal{W}_t\subseteq \mathbb{R}^n$ following a random distribution $\mathcal{D}_t$.

In \emph{operator form}, system \eqref{eq:system_state} is equivalent to
\begin{equation}
\label{eq:operator_form_state}
    \boldsymbol{\eta} = \mathbf{F}(\boldsymbol{\eta},\mathbf{u},\mathbf{x_{ref}}) + \mathbf{w}\,,\end{equation}
where $\mathbf{F}:\ell^{q}\times \ell^m \times \ell^n \rightarrow \ell^q$ is the strictly causal operator such that $\mathbf{F}(\boldsymbol{\eta},\mathbf{u},\mathbf{x_{ref}}) = (0,f_1(\eta_0,u_0,x_{ref,0}),\ldots,f_t(\eta_{t-1:0},u_{t-1:0},x_{ref,t-1:0}),\ldots)$. 

Here we have that $\xrefb \in \Xrefb\subseteq \ell^n$, where $\Xrefb$ is the set of all reference signals that the base controller can track. We also define the tracking error signal.

\begin{equation}\label{eq:error}
    \eb = \xb - \xrefb \in \ell^n.
\end{equation}
Note that $\mathbf{w}$ and $\mathbf{u}$ collects all data needed for defining the system evolution over an infinite horizon. 

\begin{definition}
 The base system (\ref{eq:operator_form_state}) asymptotically tracks references $\xrefb \in \Xrefb$ if for all $\eta_0$, for all $\wb \in \ell_p$ and for $\ub = \mathbf{0}$, it holds that $\eb \in \ell_p$ as defined in \eqref{eq:error}. 
\end{definition}

To augment the behavior of the base system~\eqref{eq:operator_form_state}, we consider nonlinear, state-feedback, time-varying control policies
\begin{align}\label{eq:control_state_2}
    \mathbf{u} = \mathbf{K}(\boldsymbol{\eta}, \mathbf{x_{ref}}) = (K_0(\eta_0, x_{ref,0}),K_1(\eta_{1:0},x_{ref,1:0}),\ldots, \nonumber \\ 
    K_t(\eta_{t:0},x_{ref,t:0}),\ldots)\,,
\end{align}
where $\mathbf{K}:\ell^q \times \ell^n \ \rightarrow \ell^m$  is a \emph{causal} operator to be designed.  Note that the controller $\mathbf{K}$ can be dynamic, as $K_t$ can depend on the entire past history of the system state.  Since for each 
$\mathbf{w} \in \ell^q$, $\mathbf{u} \in \ell^m$ and $\xrefb \in \ell^n$ the system~\eqref{eq:system_state} produces a unique state sequence $\etab \in \ell^q$, equation \eqref{eq:operator_form_state} defines the well-posed transition operator $\mathbfcal{F}:(\mathbf{u},\mathbf{w}, \mathbf{x_{ref}})\mapsto \boldsymbol{\eta}\,,$ which provides an input-to-state model of system~\eqref{eq:system_state}. 
Similarly, for each $\wb \in \ell^q$, $\xrefb \in \Xrefb$ the closed-loop system \eqref{eq:system_state}-\eqref{eq:control_state_2} produces unique trajectories. 
Hence, the closed-loop mapping $(\wb,\xrefb) \mapsto(\boldsymbol{\eta},\mathbf{u})$ is well-defined. 
Specifically, for a system $\mathbf{F}$ and a controller $\mathbf{K}$, we denote the corresponding induced closed-loop operators $(\wb,\xrefb) \mapsto\etab$ and $(\wb,\xrefb) \mapsto\mathbf{u}$ as $\Phie{F}{K}$ and $\Phiu{F}{K}$, respectively. 
Therefore, we have $\etab = \Phie{F}{K}(\wb,\xrefb)$ and $\mathbf{u} = \Phiu{F}{K}(\wb,\xrefb)$ for all $\mathbf{w} \in \ell^q$, $\xrefb \in \Xrefb$. Similarly, $\eb = \Phier{F}{K}(\wb,\xrefb)$, where $\Phier{F}{K}$ is the operator $(\wb,\xrefb) \mapsto\eb$.

\begin{definition}
The closed-loop system \eqref{eq:operator_form_state}-\eqref{eq:control_state_2} achieves reference tracking if for all $\xrefb\in \Xrefb$ and $\wb \in \ell_p$,  $\Phiu{F}{K}(\wb,\xrefb) \in \ell_p$ and $\Phier{F}{K}(\wb,\xrefb) \in \ell_p$. 
\end{definition}

Note that we do not require $\Phiu{F}{K}$ and $\Phier{F}{K}$ to be $\mathcal{L}_p$ operators. Since they take $\xrefb$ as input, which may not be an $\ell_p$ signal and even diverging (e.g., when tracking a ramp), only imposing them to be in $\mathcal{L}_p$ would not guarantee that their outputs are in $\ell_p$. Instead, we require that the outputs of $\Phiu{F}{K}$ and $\Phier{F}{K}$ are $\ell_p$ signals for all $\wb \in \ell_p$ and arbitrary $\xrefb\in\Xrefb$.

Our goal is to synthesize a control policy $\mathbf{K}$ optimizing a given cost over a finite time horizon $T$. To this aim, we consider the truncated reference signal $\mathbf{x_{ref,T:0}}$ with support  $\mathbf{X_{ref,T:0}}$ and assumes it follows a probability distribution $\mathcal{X}$.

\begin{problem}
\label{prob:boosting}
	Assume that for any $\mathbf{x_{ref,T:0}} \in \mathbf{X_{ref,T:0}}$, for any $(\wb,\ub) \in \ell_p  $ the operator $\Effe$ is such that $\eb \in \ell_p$. Find $\mathbf{K}$ solving the finite-horizon Nonlinear Optimal Control (NOC) problem:

 \begin{subequations}
    \label{NOC:cost_and_stab}
	\begin{alignat}{3}
	&\min_{\mathbf{K}(\cdot)}&& \qquad \mathbb{E}_{x_{ref,T:0}}\,\mathbb{E}_{w_{T:0}}\left[L(\eta_{T:0},u_{T:0}, x_{ref,T:0})\right] \label{NOC:cost}\\
	&\operatorname{s.t.}~~ && \eta_t = f_t(\eta_{t-1:0},u_{t-1:0},x_{ref,t-1:0})+ w_t\,, ~~ w_0 =(x_0,0)\,, \nonumber\\
	&~~&&u_t = K_t(\eta_{t:0},x_{ref,t:0})\,,~~\forall t =0,1,\ldots\,,\nonumber \\
	&~~&&\Phier{F}{K}(\wb,\xrefb),\Phiu{F}{K}(\wb,\xrefb) \in \ell_{p}\, \, \nonumber\\ 
    &~~&&\phantom{\Phier{F}{K}(\wb,\xrefb),\Phiu{F}{K}dd}\forall \wb \in \ell_p,\xrefb\in\Xrefb, \label{constraint}
	\end{alignat}
 \end{subequations}
where $L(\cdot)$ defines any \textcolor{black}{piecewise differentiable lower bounded} loss over realized trajectories $\eta_{T:0}$, input $u_{T:0}$ and reference $x_{ref,T:0}$. The expectations $\mathbb{E}_{w_{T:0}}[\cdot]$ and $\mathbb{E}_{x_{ref,T:0}}[\cdot]$ remove the effect of disturbance $w_{T:0}$ and reference $x_{ref,T:0}$ on the realized values of the loss. 
\end{problem}

In \eqref{NOC:cost}, we take the expectation over the reference. It is not, however, a binding choice. Any other costs that removes the dependence on the specific reference can be used as well, for instance $\max_{x_{ref,T:0} \in \mathcal{X}_{ref,T:0}}[\cdot]$.

\section{Main results}

\begin{figure}[h]
    \centering
    \includegraphics[width=0.9\linewidth]{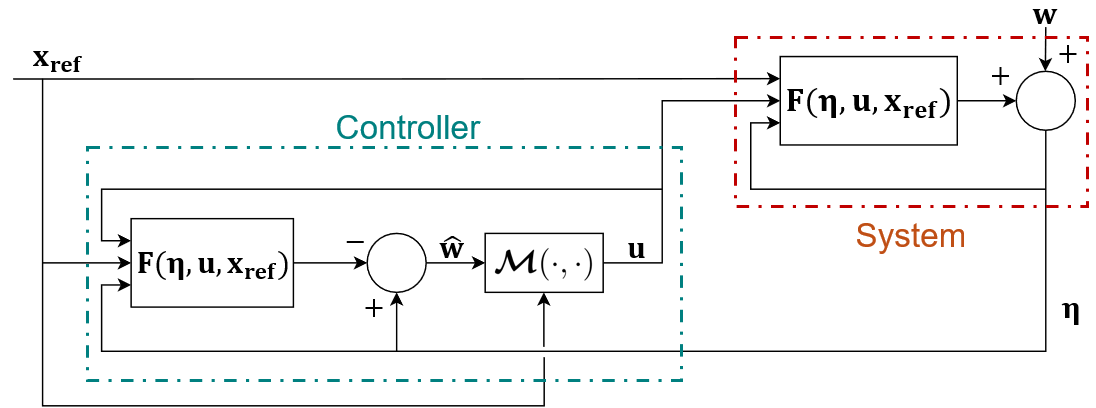}
    \caption{IMC architecture parametrizing all reference tracking controllers in terms of one operator $\Emme$}
    \label{fig:IMC Block}
\end{figure}

We show that if the base system (\ref{eq:operator_form_state}) can track any reference in a set $\Xrefb$, then, under an IMC control architecture, all and only controllers capable to track the same references can be parametrized in terms of an operator generating $\ell_p$ signals when $\wb \in  \ell_p$. The IMC control architecture includes a copy of the system dynamics, which is used for computing the estimate $\widehat \wb = \etab-\mathbf{F}(\etab,\mathbf{u},\xrefb)$ of the disturbance $\wb$. The block diagram of the proposed control architecture is represented in Figure \ref{fig:IMC Block}. We can now introduce the main result.

\begin{theorem}
	\label{th:result_IMC} 
	Assume that for any $\xrefb \in \Xrefb $ the operator $\Effe$ is such that $\eb$ defined in \eqref{eq:error} is an $\ell_p$ signal if $(\wb,\ub) \in \ell_p $ and consider the evolution of \eqref{eq:operator_form_state} where $\mathbf{u}$ is chosen as
 \begin{equation}
 \label{eq:input_M}
     \mathbf{u}=\Emme(\etab-\mathbf{F}(\etab,\mathbf{u},\xrefb),\xrefb)\,,
 \end{equation} 
 for a causal operator $\Emme:\ell^q\times \ell^n \rightarrow \ell^m$. 
 Let $\mathbf{K}$ be the operator such that $\mathbf{u}=\mathbf{K}(\etab,\xrefb)$ is equivalent to \eqref{eq:input_M}.\footnote{This operator always exists because $\mathbf{F}(\mathbf{\etab},\mathbf{u}, \xrefb)$ is strictly causal. Hence $u_t$ depends on the inputs $u_{t-1:0}$ and can be computed recursively from past inputs, $\eta_{t:0}$ and $x_{ref,t:0}$ --- see formula \eqref{eq:controller_IMC}.} %
 
 The following statements hold true.
\begin{enumerate}
		\item If $\Emme(\etab-\mathbf{F}(\etab,\mathbf{u},\xrefb),\xrefb) \in \ell_{p} \,$, then for any $\wb \in \ell_p$ and $\xrefb \in \Xrefb$, the closed-loop system is such that $\eb\in \ell_p$ and $\ub\in \ell_p$  
		\item For any causal policy $\bmC$ such that  $\Phier{\Fb}{\bmC}(\wb,\xrefb) \in \ell_p  $ and $~\Phiu{\Fb}{\bmC}(\wb,\xrefb) \in \ell_p \,$, for any $ \wb \in \ell_p$ and  $\xrefb \in \Xrefb$, the operator
			\begin{equation}
		\label{eq:choice_Emme}
		\Emme=\Phiu{\Fb}{\bmC}\,, 
		\end{equation}
		 gives $\bmK=\bmC$. 
	\end{enumerate}
\end{theorem}

The proof of Theorem \ref{th:result_IMC} can be found in Appendix \ref{proof1}. Note that in the statement, $\etab-\mathbf{F}(\etab,\mathbf{u},\xrefb)$ is the estimation of $\wb$ done internally by the controller by using the variables available to it. As there is no model mismatch, this estimation is exact. For a chosen operator $\Emme$, the control input is simply computed as:
\begin{subequations}
\label{eq:controller_IMC}
\begin{align}
    &\widehat{w}_t = \eta_t - f_t(\eta_{t-1:0},u_{t-1:0}, x_{ref,t-1:0})\,, \label{eq:controller_IMC_1}\\
    &u_t = \mathcal{M}_t(\widehat{w}_{t:0},x_{ref,t:0})\,.\label{eq:controller_IMC_2}
\end{align}
\end{subequations}

\begin{figure}[h]
    \centering
    \includegraphics[width=0.7\linewidth]{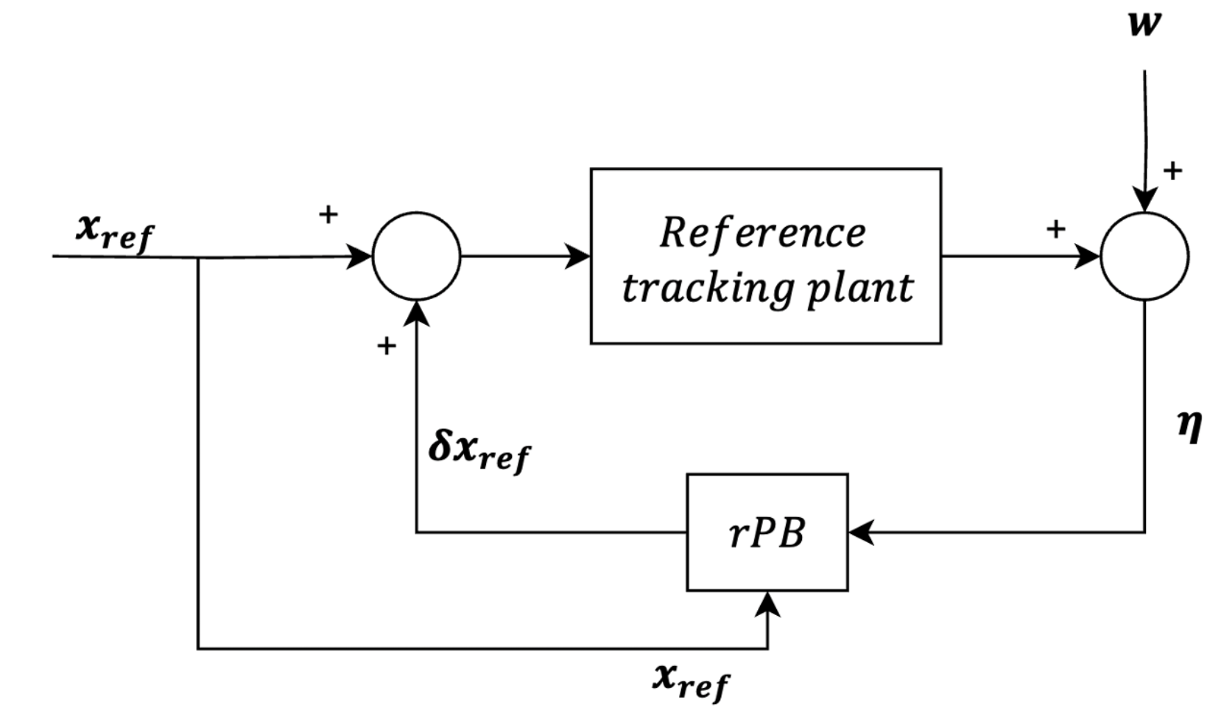}
    \caption{Diagram of the rPB with reference governor architecture acting on the reference signal}
    \label{fig:diagram_rpb}
\end{figure}

An intuitive way to understand rPB is to see it as a reference governor \cite{kolmanovsky2014reference}, which consists in considering the subcase where $\mathbf{F}(\boldsymbol{\eta},\mathbf{u},\mathbf{x_{ref}})=\mathbf{F}(\boldsymbol{\eta},\mathbf{u}+\mathbf{x_{ref}})$.  In this setting, we rename the rPB output $\ub$ as $\delta \xrefb$, to emphasize how it can be seen as an offset to the target, inducing new behavior in the base system. Since $\Emme(\etab,\xrefb)\in \ell_p$, this offset vanishes asymptotically, ensuring that the tracking property of the system is preserved. A diagram of the closed-loop system with rPB in reference governor structure is shown in Figure \ref{fig:diagram_rpb}. 

\begin{remark} \label{remark}
    Even if $\Emme$ takes as input $\xrefb\in \ell^n$ which does not vanish as $t \rightarrow +\infty$, it is possible to ensure that its output always remains in $\ell_p$. Following the ideas proposed in \cite{L2operator}, a necessary and sufficient condition for this to hold is to design the operator $\Emme$ as the product of two operators: 

\begin{equation} \label{eq:newM}
    \Emme(\wb,\xrefb) = \Emme_1(\wb)*\Emme_2(\wb,\xrefb), 
\end{equation}
where $\Emme_1(\wb) \in \ell_p$ and $\Emme_2(\wb,\xrefb) \in \ell_\infty$, $\forall \wb \in \ell_p, \xrefb \in \Xrefb$. The boundedness on $\Emme_2$ ensures that $\Emme(\wb,\xrefb) \in \ell_p$. This condition can be enforced, for example by taking $\Emme_2$ as a  multilayer perceptron (MLP) neural network with a sigmoid output layer activation function. 

\end{remark}

\subsection{Model Mismatch}\label{Model Mismatch}

We now consider the case where the internal model in the IMC controller is not a perfect reconstruction of the plant. We only consider achievable reference trajectories. A reference trajectory is achievable if the base system can perfectly track it in the absence of external disturbances.

\begin{definition}
    A trajectory $\xrefb \in \Xrefb$ is achievable if there exists $ \etab = (\xrefb,\vb)^\top$ such that $\etab = \mathbf{F}(\boldsymbol{\eta},\mathbf{0},\mathbf{x_{ref}})$. The set of all achievable trajectories is denoted as $\Xrefba$.
\end{definition}
We show that if the mismatch can be modeled as an i.f.g $\ell_p$-stable operator, then it is possible to design $\Emme$ with a sufficiently small incremental gain such that steady-state reference tracking is preserved in closed loop. 

Let us denote the nominal model available for design as ${\hatbf{F}}(\etab,\ub, \xrefb)$ and the real unknown plant as 
\begin{equation}
    \label{eq:mismatch_plant}
    \mathbf{F}(\etab,\ub, \xrefb) = \hatbf{F}(\etab,\ub, \xrefb) + \bm{\Delta}(\bmx,\bmu,\xrefb)\,,
\end{equation}
where $\bm{\Delta}$ is a strictly causal operator representing the model mismatch. 
%where the operator $\bm{\Delta}$ is the model mismatch. 
Let $\delta_t(x_{t-1:0},u_{t-1:0},x_{ref,t-1:0})$ be the time representation of the mismatch operator $ \mathbf{\Delta}$. Since for each sequence of disturbances $\mathbf{w} \in \ell^q$, inputs $\mathbf{u} \in \ell^m$ and reference $\xrefb \in \Xrefb$ the dynamics represented by \eqref{eq:system_state} with $f_t(\eta_{t-1:0},u_{t-1:0},x_{ref,t-1:0})$ replaced by $\widehat{f}_t(\eta_{t-1:0},u_{t-1:0},x_{ref,t-1:0})+\delta_t(x_{t-1:0},u_{t-1:0})$  produces a unique state sequence $\etab \in \ell^q$, the equation 
    \begin{equation}
\label{eq:operator_form_perturbed}
    \etab = \mathbf{F}(\etab,\mathbf{u},\xrefb) + \mathbf{w}\,,
\end{equation}
defines again a unique transition operator $\Effe:(\mathbf{u},\mathbf{w},\xrefb)\mapsto \boldsymbol{\eta}$, which provides an input-to-state model of the perturbed system. Similarly, the unique transition operator $\Effe^x:(\mathbf{u},\mathbf{w},\xrefb)\mapsto \bmx$ can be defined, providing an input-to-plant-state map. 

Letting $\alpha_{\bm{\Delta}}$ be the upper bound of the incremental  $\mathcal{L}_p$-gain of the model mismatch $\bm{\Delta}$, we show that it is possible to design controllers $\mathbf{K}$ that comply with the following robust reference tracking constraints:
\begin{align}
\label{eq:robust_stability}
    &(\bm{\Phi}^{*}[\widehat{\bmF}+\bm{\Delta},\mathbf{K}](\wb,\xrefb)) \in \ell_{p}\,, ~ 
    *\in\{\bme,\bmu \} \,, ~
     \nonumber \\
    & \phantom{(\bm{\Phi}^{*}[\widehat{\bmF}+\bm{\Delta},\mathbf{K}]) }\forall \xrefb\in \Xrefba ,\forall \bm{\Delta}|~\alpha(\bm{\Delta}) \leq \alpha_{\bm{\Delta}}.\,
\end{align}

\begin{theorem}
	\label{th:result_robust} 
	Assume that the mismatch operator $\bm{\Delta}$ in \eqref{eq:mismatch_plant} has finite incremental $\mathcal{L}_p$-gain $\alpha(\mathbf{\Delta})$. Furthermore, assume that
 the operator $\Effe^x$  has a finite incremental $\mathcal{L}_p$-gain $\alpha(\mathbf{\Effe}^x)$.  Then, for any $\Emme$ such that 
 \begin{equation}
 \label{eq:condition_robustness}
 \alpha(\bmcalM)<\alpha(\mathbf{\Delta})^{-1}(\alpha(\Effe^x)+1)^{-1}\,,
 \end{equation}
 the control policy given by
 \begin{subequations}
 \label{eq:robust_policy}
 \begin{align}
 &\widehat{w}_t = \eta_t - \widehat{f}_t(\eta_{t-1:0},u_{t-1:0},x_{ref,t-1:0})\,,\label{eq:robust_policy_1}\\
 &u_t = \mathcal{M}_t(\widehat{w}_{t:0},x_{ref,t-1:0})\,,\label{eq:robust_policy_2}
 \end{align}
 \end{subequations}
 ensures that the closed loop maps verifies $\Phiu{F}{K}(\wb,\xrefb) \in \ell_p$ and $\Phier{F}{K}(\wb,\xrefb) \in \ell_p $ for all $\xrefb\in\Xrefb$ and $\wb\in\ell_p$.
 \end{theorem}

\noindent The proof can be found in Appendix \ref{proof2}.

\subsection{Implementation of the  operator $\mathcal{M}$}

The results derived in the previous sections allow us to get rid of constraint (\ref{constraint}) in the NOC problem, as long as we optimize over operators $\Emme$ designed as in Remark \ref{remark}. Without system model mismatch, optimizing over operators $\Emme_1 \in \mathcal{L}_p$ and $\Emme_2$ such that $\Emme_2(\wb,\xrefb) \in \ell_\infty$ guarantees that closed-loop reference tracking is preserved. 

However, optimizing directly over such $\Emme$ is an infinite dimensional problem, so we instead chose $\Emme_1$ from a family of parametrized $\mathcal{L}_p$-stable operators. These can be optimized over a finite number of parameters, making the problem tractable. Furthermore, we consider dynamical models offering free parametrization of these operators, ensuring that the NOC problem can be solved using unconstrained optimization. $\Emme_2$ is parametrized as an MLP, which also has a finite number of parameters.

Several dynamical models with these properties exists. In this paper, we will use contractive RENs \cite[Eq.~(15)]{REN}  because they offer free parametrization of both $\mathcal{L}_p$ and i.f.g $\ell_p$-stable operators. Other options include classes of SSMs \cite{forgione2021dynonet}, port-Hamiltonian based neural networks \cite{zakwan2024neural}, or Lipchitz-bounded deep neural networks \cite{wang2023direct}.   

By using the direct parametrization proposed in Section 5 of \cite{REN} for $\Emme_1$, it is possible to reformulate (\ref{prob:boosting}) as a unconstrained optimization problem. The parameters to optimize over become $\theta \in \mathbb{R}^d$, which includes the parameters of both $\Emme_1$ and $\Emme_2$. We also replace the intractable exact average by an empirical approximation obtained using a set of samples $\{(w_{T:0}^s,x_{ref,T:0}^s)\}_{s=1}^S$ drawn from the distributions $\mathcal{D}_{T:0}$ and $\mathcal{X}$. Problem (\ref{prob:boosting}) then becomes:  

\begin{problem}{Finite horizon unconstrained problem}
\label{prob:boosting2}
 \begin{subequations}
    \label{NOC2:cost_and_stab_free}
	\begin{alignat}{3}
	&\min_{\theta \in \mathbb{R}^d}&& \qquad \frac{1}{S} \sum_{s=1}^SL(\eta^s_{T:0},u^s_{T:0}, x^s_{ref,T:0})\label{NOC2:cost}\\
	&\operatorname{s.t.}~~ && \eta^s_t = f_t(\eta^s_{t-1:0},u^s_{t-1:0},x^s_{ref,t-1:0})+ w^s_t\,, \nonumber \\ 
    &~~&&~~\phantom{\eta^s_t = f_t(\eta^s_{t-1:0},u^s_{t-1:0},)\,\,\,} w^s_0 =(x^s_0,0)\,,\label{Model} \\
	&~~&&u^s_t = \mathcal{M}_t(\theta)(w^s_{t:0},x^s_{ref,t:0})\,,~~\forall t =0,1,\ldots\,, \label{NN}
	\end{alignat}
 \end{subequations}
 \end{problem}
where $\eta_{T:0}^s$ and $u_{T:0}^s$ are the inputs and states obtained when the disturbance $w_{T:0}^s$ and reference $x_{ref,T:0}^s$ is applied. As we roll out \eqref{Model} and \eqref{NN} over the horizon, the parameters we optimize appear multiple times within each trajectory. The absence of constraints on $\theta$ allows us to leverage powerful optimization frameworks such as PyTorch \cite{paszke2019pytorch}, using a backpropagation-through-time approach \cite{werbos1990backpropagation} to design the rPB controller efficiently.

\section{Simulation examples}\label{sec:numerical}

This section shows the application of the rPB framework to cooperative robotics problems. The new approach is first compared to the standard PB framework in a simpler setup, followed by a demonstration of its potential in a more complex scenario. 

In both cases we consider two point-mass robots, each with position $p_t^{[i]} \in \mathbb{R}^2$ and velocity $q_t^{[i]} \in \mathbb{R}^2$, for $i=1,2$, subject to nonlinear drag forces (e.g., air or water resistance) and obeying to the dynamics described in \cite{10633771}. 

Moreover, unlike \cite{10633771}, the robots are equipped with an integral controller for tracking any constant set-point. However, this integral controller cannot provide any additional desired behaviors besides tracking, like collision and obstacle avoidance. 
We use rPB to improve on this. Similarly to Figure \ref{fig:diagram_rpb}, we design the rPB to act as an offset $\delta_{ref} \in \mathbb{R}^4$ that influences all the set points of the system. The integral controller becomes 
\begin{align*}
    & {F}_{t}^{[i]} = {K_v}^{[i]}v^{[i]}_t + {K_p}^{[i]} p_{t}^{[i]}\\
    & v_{t+1}^{[i]} = v_t^{[i]} + ((\bar p^{[i]}+\delta_{ref}^{[i]})-p_{t}^{[i]}), 
\end{align*}
for $i =1,2$, where  $v^{[i]}_t \in \mathbb{R}^2$ is the state of the integrator of robot $i$, $\bar p^{[i]}\in \mathbb{R}^2$ is the target, and $K_v^{[i]} = \operatorname{diag}(k_{v,1}^{[i]},k_{v,2}^{[i]}) \in \mathbb{R}^{2 \times 2}$ and $K_p^{[i]} = \operatorname{diag}(k_{p,1}^{[i]},k_{p,2}^{[i]}) \in \mathbb{R}^{2 \times 2}$ are the gains of the base controller.

Intuitively, rPB can be understood as dynamically shifting the target that the base controller tracks to shape the system's transient response. If the actual target is obstructed by an obstacle, rPB can temporarily offset it to the side, allowing the robot to bypass the obstacle. Once past the obstruction, the target seamlessly returns to its original position in steady state, ensured by the $\mathcal{L}_p$ properties of the controller.

We use a loss which penalizes the distance to the target, as well as collisions between robots and with the obstacle. The specific loss formulation can be found in Appendix C of \cite{kirsch2025boostingtransientperformancereference}. In the following experiments, we train using stochastic gradient descent with Adam, setting a learning rate of $1\times 10^{-4}$. The REN we used had internal linear and nonlinear states of twelve dimensions and the MLP comprised four linear layers with widths 15, 20, 14, and output dimension, with sigmoid activations after each hidden layer.

\begin{comment}
   \begin{figure}[h]
    \centering
    \begin{subfigure}[b]{0.4\linewidth}
        \centering
        \includegraphics[width = \linewidth]{figures/training_PB.png}
        \caption{Training data for bPB}
        \label{fig:range_cross}
    \end{subfigure}
    \begin{subfigure}[b]{0.4\linewidth}
        \centering
        \includegraphics[width = \linewidth]{figures/training_rPB.png}
        \caption{Training data for rPB}
        \label{fig:ll}
    \end{subfigure}
    \caption{Training data for bPB and rPB with one example rollout. The dots represent the initial position data and the stars/crosses the reference data. bPB can only train on one reference.  }
    \label{fig:training_data}
\end{figure} 
\end{comment}

\subsection{Benchmark against original PB} \label{Benchmark}

First, we compare the previously existing version of PB (referred to here as bPB) with rPB, on the task of passing through a tight corridor. Both experiments start with robot 1 (blue) in position $p^{[1]}_{0} = (-2,-2)$ and robot 2 (orange) in position $p^{[2]}_{0} = (2,-2)$. bPB was trained with one target for each robot: $p^{[1]}_{ref}=(2,2)$ and $p^{[2]}_{ref} = (-2,2)$. Since the reference is an input for rPB, it was possible to train it with any pairs of references $(p^{[1]}_{ref},p^{[2]}_{ref}) \in \mathcal{P}$, where $\mathcal{P} = \big\{(p^{[1]}_{ref},p^{[2]}_{ref}) \in \mathbb{R}^4 \,\,| \,\,||p^{[1]}_{ref}-p^{[2]}_{ref}||_2 \geq 1, -2 \leq p^{[i]}_{ref,x} \leq 2, p^{[i]}_{ref,y} = 2, i = 1,2  \big\}$. This corresponds to a set of targets above the obstacle far enough one from the other to not cause robot collision. For each training session, a gaussian noise with standard deviation of 0.5 was added to the initial position. 
%The training data can been seen in figure \ref{fig:training_data}. 
bPB was trained on 30 rollouts for 300 epochs and rPB on 200 rollouts for the same number of epochs.

\begin{figure}[h]
    \centering
    \includegraphics[width=0.8\linewidth]{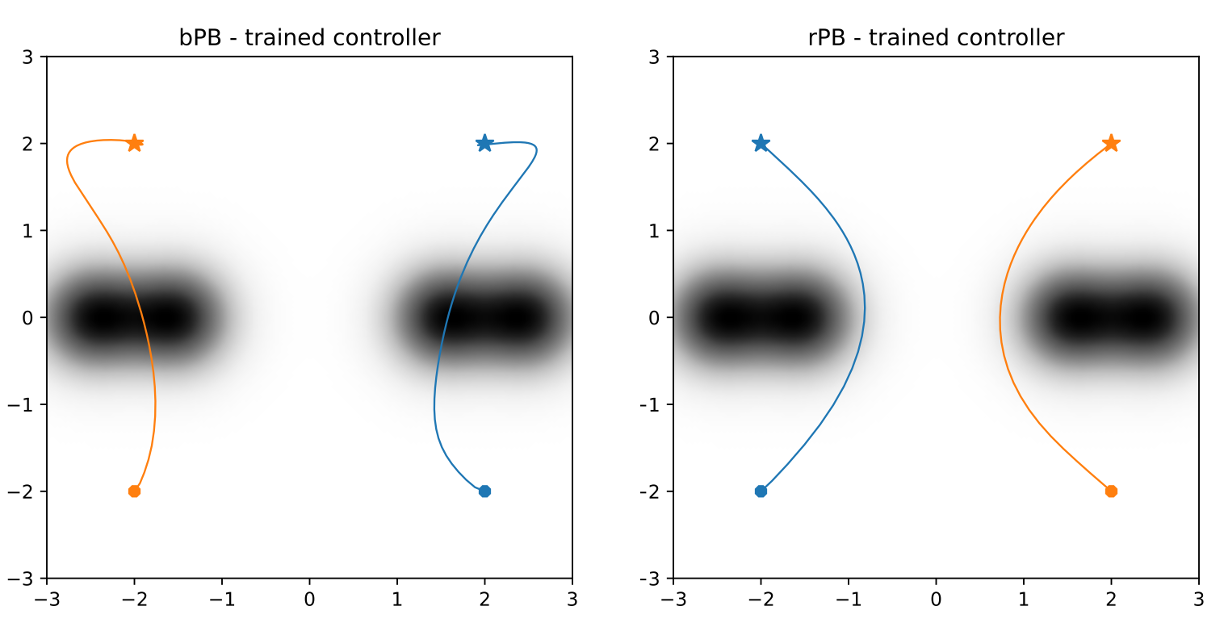}
    \caption{Closed-loop trajectories for bPB and rPB controllers after training}
    \label{fig:benchmark}
\end{figure}

The closed-loop trajectories for both bPB and rPB are shown in Figure \ref{fig:benchmark} for targets $p^{[1]}_{ref}=(-2,2)$ and $p^{[2]}_{ref} = (2,2)$. Note that this target is different than the one bPB has been trained on. For bPB, the resulting trajectories lead to important collisions with the obstacles. The robots still reach the target due to the $\mathcal{L}_p$ nature of the PB controller and because the base controller is designed to ensure this. On the other hand, rPB generates trajectories that appear to be the shortest while avoiding collisions with the obstacles.  

\subsection{Mountain range example} \label{Mountain range}

We now apply rPB to a more complex task. Once again, rPB is applied to the same two-robot system, but this time in a different environment. Instead of navigating a tight corridor, the robots must traverse an array of obstacles. Depending on their initial conditions and targets, they have multiple possible paths to reach their goal, as they can maneuver through various gaps between obstacles. We train the controller for 600 epochs on 2000 rollouts.

\begin{figure}[h]
    \centering
    \begin{subfigure}[b]{0.3\linewidth}
        \centering
        \includegraphics[width = \linewidth]{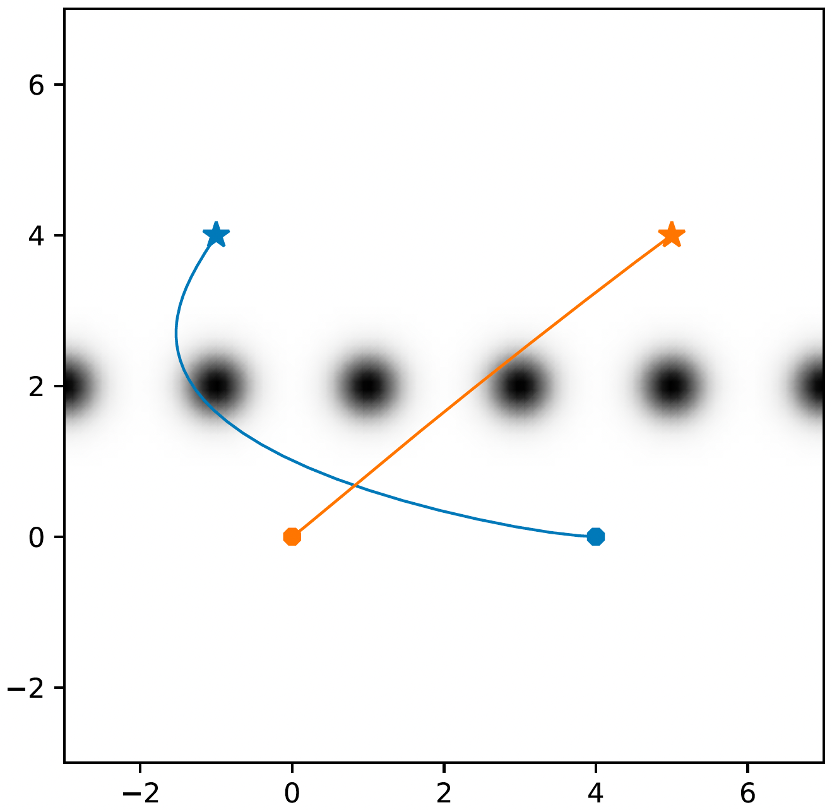}
        \caption{Base controller - Diagonal targets}
        \label{fig:range_cross_base}
    \end{subfigure}
    \begin{subfigure}[b]{0.3\linewidth}
        \centering
        \includegraphics[width = \linewidth]{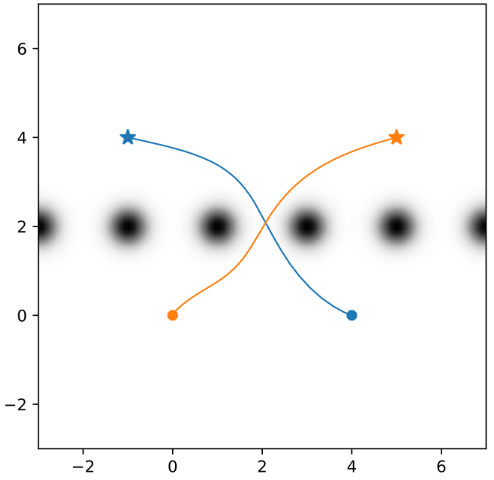}
        \caption{rPB - Diagonal targets}
        \label{fig:range_cross}
    \end{subfigure}
    \begin{subfigure}[b]{0.3\linewidth}
        \centering
        \includegraphics[width = \linewidth]{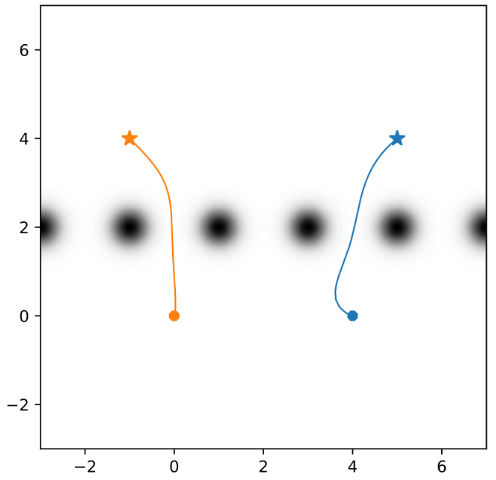}
        \caption{rPB - Straight targets}
        \label{fig:range_straight}
    \end{subfigure}
    \caption{Closed loop trajectories with and without the trained rPB controller. The full animated trajectories can be found at \cite{kirsch_gifs_rPB_2025}}
    \label{fig:mountain_range}
\end{figure}

Figure \ref{fig:mountain_range} shows two test closed-loop trajectories resulting from a single training process as well as a trajectory with only the base controller. Comparing Figure \ref{fig:range_cross_base} and \ref{fig:range_cross} clearly underlines the transient improvement brought by rPB as it avoids obstacle collisions. Furthermore, no collisions occurred between the robots using rPB across the 500 test scenarios, which varied in both their initial conditions and their targets. Additionally, in Figure \ref{fig:range_cross} and Figure \ref{fig:range_straight}, the robots choose the gap that minimizes the path length to the target. The trained controller thus enhances the transient performance of the system. 

\section{Conclusions}

We have introduced the rPB framework, providing a parametrization of all and only reference tracking controllers in terms of an  operator with mild conditions on its input-to-output behavior. Furthermore, we have shown that our approach can be robust to mismatches between the real plant and its nominal model. The effectiveness of our method was demonstrated on a robotic system, where the rPB controller successfully optimized transient behavior in a complex environment while preserving reference tracking. Notably, the controller required only a single training phase for a wide range of targets, representing a significant improvement over the standard PB framework.

Future research will consider the extension of rPB to distributed control architecture, or to systems with input-to-output representations.

\appendix
\subsection{Proof of Theorem 1}\label{proof1}

We prove statement $1)$ of the theorem.  For compactness, we define $\hatbf{w}=\etab -\mathbf{F}(\boldsymbol{\eta},\mathbf{u},\mathbf{x_{ref}})$. 
    Since there is no model mismatch between the plant $\Effe$ and the model $\mathbf{F}$ used to define $\hatbf{w}$, one has $\hatbf{w}=\wb$, hence opening the loop. 
	Therefore, by the definition of the closed-loop maps,  one has $\Phiu{\Fb}{ \bmK}=\Emme$ and $\Phie{\Fb}{\bmK}(\wb,\xrefb)=\mathbf{F}(\boldsymbol{\eta},\Emme(\wb,\xrefb),\mathbf{x_{ref}})+\wb$, $\forall \wb \in \ell_p$.
	When $\wb\in\ell_p$, one has $\Phiu{\Fb}{ \bmK}(\wb,\xrefb)\in\ell_p$ because $\Emme(\wb,\xrefb)\in\ell_p$. 
    Moreover, given that  $\Emme(\wb,\xrefb)\in\ell_p$ and $\Effe$ is such that $\eb\in \ell_p $ when $(\wb,\ub) \in \ell_p$, the operator $(\wb,\xrefb)\mapsto \etab$ defined by the composition of the operators $(\wb,\xrefb)\mapsto(\Emme(\wb,\xrefb),\wb,\xrefb)$ and $\Effe$ is also such that $\eb\in \ell_p $ when $(\wb,\ub) \in \ell_p$.  
	 We prove $2)$.   
  Set, for short, 
  $\Psie=\Phie{\Fb}{\bmC}$, 
  $\Psiu=\Phiu{\Fb}{\bmC}$, 
  $\Phieno =\Phie{\Fb}{ \bmK}$, and 
  $\Phiuno=\Phiu{\Fb}{ \bmK}$. By assumption, one has $\Emme=\Psiu$ and since $\Psiu(\wb,\xrefb)\in\ell_p$ also $\Emme(\wb,\xrefb)\in\ell_p$. By definition, $\Phiuno$ is the operator $(\wb,\xrefb)\mapsto\bmu$ and, as $\hatbf{w}=\wb$, it coincides with $\Emme$. Hence \begin{equation}
\label{eq:PhiuisPsiu}
\Psiu=\Phiuno\,.
\end{equation}
It remains to prove that $\Phieno = \Psie$. We proceed by induction. 
First, we show that $\Psi^\eta_{0}=\Upsilon^\eta_{0}$, where $\Psi^\eta_{0}$ and $\Upsilon^\eta_{0}$ are the components of $\Psie$ and $\boldsymbol{\Upsilon}^{\boldsymbol{\eta}}$ at time zero.
Since $f_0=0$ and $w_0=(x_0^\top,0_v^\top)^\top$, one has from \eqref{eq:system_state} that the closed-loop map $w_0\mapsto x_0$ is the identity, irrespectively of the controller. Furthermore, the maps $w_0\mapsto v_0$, $x_{ref,0}\mapsto x_0$ and $x_{ref,0}\mapsto v_0$ are all 0. Therefore 

\begin{equation}
    \Upsilon_0^\eta = \Psi_0^\eta = \begin{pmatrix}
I & 0 \\
0 & 0
\end{pmatrix}.
\end{equation}  
Assume now that, for a positive $j \in \mathbb{N}$ we have $\Upsilon^\eta_i = \Psi^\eta_i$ for all $0\leq i \leq j$. 
Since $(\Phieno,\Phiuno)$ and $(\Psie, \Psiu)$ are closed-loop maps, from \eqref{eq:operator_form_state} they verify
	\begin{align}
	\label{eq:achievability_proof}
	\Upsilon^\eta_{j\hspace{-0.04cm}+\hspace{-0.04cm}1} \hspace{-0.1cm}=\hspace{-0.1cm} F_{j\hspace{-0.04cm}+\hspace{-0.04cm}1}(\Upsilon^\eta_{j:0},\Upsilon^u_{j:0})\hspace{-0.04cm}+\hspace{-0.01cm}\begin{pmatrix}
I & 0 \\
0 & 0
\end{pmatrix},\\
\Psi^\eta_{j\hspace{-0.01cm}+\hspace{-0.04cm}1} \hspace{-0.1cm}=\hspace{-0.1cm} F_{j\hspace{-0.04cm}+\hspace{-0.01cm}1}(\Psi^\eta_{j:0},\Psi^u_{j:0})\hspace{-0.01cm}+\hspace{-0.04cm}\begin{pmatrix}
I & 0 \\
0 & 0
\end{pmatrix}.
\end{align}
But, from \eqref{eq:PhiuisPsiu}, one has $\Psi^u_{j:0}=\Upsilon^u_{j:0}$ and, by using the inductive assumption, one obtains $\Upsilon^\eta_{j\hspace{-0.04cm}+\hspace{-0.04cm}1}=\Psi^\eta_{j\hspace{-0.04cm}+\hspace{-0.04cm}1}$. So $\mathbf{K=C}$. \qed
\subsection{Proof of Theorem 2}\label{proof2}

We first show that operators $\mathbf{F}$ and $\Effe$ verify $\mathbf{F}(\Effe(\mathbf{u},\mathbf{w}),\mathbf{u},\xrefb)=\Effe(\mathbf{u},\mathbf{w},\xrefb)-\mathbf{w}\,.$
\begin{comment}
   \begin{equation}
     \label{eq_F_adn_Effe}
     \mathbf{F}(\Effe(\mathbf{u},\mathbf{w}),\mathbf{u},\xrefb)=\Effe(\mathbf{u},\mathbf{w},\xrefb)-\mathbf{w}\,.
 \end{equation}  
\end{comment}
This follows by substituting $ \etab=\Effe(\mathbf{u},\mathbf{w}, \xrefb)$  in \eqref{eq:operator_form_perturbed}. We now compute the incremental $\mathcal{L}_p$-gain of the map: $(\mathbf{u},\mathbf{w},\xrefb)\mapsto\hatbf{w}$, linking the system inputs to the estimate of the disturbance:
 \begin{align}
    \hatbf{w} &= \Effe(\mathbf{u},\mathbf{w},\xrefb)-\hatbf{F}(\Effe(\mathbf{u},\mathbf{w},\xrefb),\mathbf{u},\xrefb) \nonumber \\
    &=\mathbf{F}(\Effe(\mathbf{u},\mathbf{w},\xrefb),\mathbf{u},\xrefb)- \hatbf{F}(\Effe(\mathbf{u},\mathbf{w},\xrefb),\mathbf{u},\xrefb)\nonumber \\
    & \qquad\qquad\qquad\qquad\qquad\qquad\qquad\qquad\qquad\qquad\qquad+\mathbf{w} \nonumber\\
    &= \mathbf{\Delta}(\Effe^x(\mathbf{u},\mathbf{w},\xrefb),\mathbf{u},\xrefb)+ \mathbf{w} \,,\label{eq:u_to_twDelta}
\end{align} 
 
 Using the definition of incremental $\mathcal{L}_p$-gain for the operator $\mathbf{y}=\mathbf{\Delta}(\mathbf{x},\mathbf{u},\xrefb)$ one has 
 $||\mathbf{y_1}-\mathbf{y_2}||_p \leq \alpha(\mathbf{\Delta})(||\mathbf{x_1}-\mathbf{x_2}||_p+||\mathbf{u_1}-\mathbf{u_2}||_p+||\xrefb_1-\xrefb_2||_p)$, for any input pairs. In our case, we consider one arbitrary trajectory with inputs $\mathbf{w_1}=\wb$ and $\mathbf{x_{ref,1}}=\xrefb$ resulting in the plant state $\mathbf{x_1} = \xb$ and the input $\mathbf{u_1} =\ub$. The second trajectory we consider perfectly tracks the same reference with no disturbances so $\mathbf{x_{ref,2}}=\xrefb$ and $\wb_2=0$. For this experiment, $\mathbf{x_2}= \xrefb$ and $\mathbf{u_2}=0$ because the rPB is not active. Thanks to the assumption made on considering only achievable references, this trajectory exists for the system. The incremental $\mathcal{L}_p$ nature of the operator $\Effe^x$ thus means that regarding these two experiments: 

\begin{equation} \label{inc_l2_fx}
     ||\xb-\xrefb||_p \leq \alpha(\Effe^x)(||\ub||_p+||\wb||_p).
 \end{equation}

Assuming that $\Emme$ is also an operator with finite incremental $\mathcal{L}_p$-gain:  

\begin{equation} \label{inc_l2_M}
     ||\ub||_p \leq \alpha(\Emme)(||\hatbf{\wb}-\hatbf{\wb}_{\mathbf{ref}}||_p),
 \end{equation}
where $\hatbf{\wb}_{\mathbf{ref}}$ is the noise reconstruction in the  trajectory perfectly tracking the reference. Note that having $\wb=\mathbf{0}$ for the base system to perfectly track the reference does not necessarily mean that $\hatbf{\wb}_{\mathbf{ref}}=\mathbf{0}$ because of model mismatch. In both trajectory the second input is $\xrefb$, so it cancels out in the right hand side of the incremental $\mathcal{L}_p$ inequality.

By using \eqref{eq:u_to_twDelta}, \eqref{inc_l2_fx} and \eqref{inc_l2_M}, one obtains\footnote{For improving the clarity of the proof, from here onwards, we omit the subscript $p$ of the signal norms.}
  \begin{align*}
&||\hatbf{w}-\hatbf{\wb}_{\mathbf{ref}}||\leq \alpha(\mathbf{\Delta})(||\Effe^x(\mathbf{u},\mathbf{w},\xrefb)||+||\mathbf{u}||)+||\mathbf{w}|| \nonumber \\
&\leq \alpha(\mathbf{\Delta})(\alpha(\Effe^x)||\mathbf{w}||+\alpha(\Effe^x) ||\mathbf{u}||+||\mathbf{u}||)+||\mathbf{w}|| \nonumber \\
&\leq (\alpha(\mathbf{\Delta})\alpha(\Effe^x)+1)||\mathbf{w}||\\
&\phantom{(\alpha(\mathbf{\Delta})\alpha(\Effe^x)}+\alpha(\mathbf{\Delta})(\alpha(\Effe^x)+1) \alpha(\Emme)||\hatbf{w}-\hatbf{\wb}_{\mathbf{ref}}||\,.
\end{align*}
\textcolor{black}{By gathering all the terms involving $||\hatbf{w}-\hatbf{\wb}_{\mathbf{ref}}||$ to the left-hand side we obtain
\begin{align*}
    &(1-\alpha(\bm{\Delta}) \alpha(\Emme)\left(\alpha(\Effe^x)+1\right))||\hatbf{w}-\hatbf{\wb}_{\mathbf{ref}}||\leq \\
     &\phantom{(1-\alpha(\bm{\Delta}) \alpha(\Emme)\left(\alpha(\Effe^x)+1\right))}(\alpha(\bm{\Delta})\alpha(\Effe^x)+1)||\mathbf{w}||\,.
\end{align*}
Since \eqref{eq:condition_robustness} holds, we have $1-\alpha(\bm{\Delta})\alpha(\Emme)(\alpha(\Effe^x)+1)>0$, and hence}
\begin{equation}
\label{eq:map_w->w_hat}
    ||\hatbf{w}-\hatbf{\wb}_{\mathbf{ref}}||\leq \left(\frac{\alpha(\bm{\Delta})\alpha(\Effe^x)+1}{1-\alpha(\bm{\Delta}) \alpha(\Emme)\left(\alpha(\Effe^x)+1\right)} \right)||\mathbf{w}||\,.
\end{equation}

Next, we plug the upper bound \eqref{eq:map_w->w_hat} into the inequality $||\mathbf{u}||\leq\alpha(\Emme)||\mathbf{w}||$ to obtain
\begin{equation}
    \label{eq:robust_loop_u}
    ||\mathbf{u}||\leq \left(\frac{\alpha(\Emme)\left(\alpha(\bm{\Delta})\alpha(\Effe^x)+1\right)}{1-\alpha(\bm{\Delta}) \alpha(\Emme)(\alpha(\Effe^x)+1)} \right)||\mathbf{w}||\,,%\left(\frac{\alpha(\Emme)g\left(\alpha(\bm{\Delta})\right)}{1-g\left(\alpha(\Emme)\right)\alpha(\bm{\Delta})}\right)|\wb|\,,
\end{equation}
and subsequently, we plug \eqref{eq:robust_loop_u} into 
the inequality $||\mathbf{e}||\leq \alpha(\Effe^x)(||\mathbf{u}||+||\mathbf{w}||)$ to obtain
%the relationship $\mathbf{x}= \Effe(\mathbf{u},\mathbf{w})$ to obtain
\begin{equation}
    \label{eq:robust_loop_x}
    ||\mathbf{e}||\leq \left(\alpha(\Effe^x)\frac{1+\alpha(\Emme)\left(1-\alpha(\bm{\Delta})\right)}{1-\alpha(\bm{\Delta}) \alpha(\Emme)(\alpha(\Effe^x)+1)}\right)||\mathbf{w}||\,.
\end{equation}
The last step is to verify that the gains in \eqref{eq:robust_loop_u} and \eqref{eq:robust_loop_x} are positive values when the gain of $\Emme$ is sufficiently small. \textcolor{black}{Since  \eqref{eq:condition_robustness} holds}, the denominator in \eqref{eq:robust_loop_u} is positive. Since the numerator of \eqref{eq:robust_loop_u} is always positive, we conclude that the map $(\wb,\xrefb)\rightarrow \ub$ has an incremental $\mathcal{L}_p$-gain. Similarly for \eqref{eq:robust_loop_x}, since \eqref{eq:condition_robustness} implies that $\alpha(\Emme) \alpha(\bm{\Delta})(\alpha(\Effe^x)+1)<1$, we have that both numerator and denominator are positive. Because $\wb \in \ell_p$ this implies that both $\eb \in \ell_p$ and $\ub \in \ell_p$ in closed-loop, as desired.  \qed

\bibliographystyle{IEEEtran}
\bibliography{bibliography}

\end{document}